\documentclass[DIV20, parskip = false]{scrartcl}
\usepackage{graphicx}
\usepackage{amsmath}
\usepackage{SIunits}
\usepackage{authblk}
\usepackage{cite}
\usepackage{color}
\usepackage{ulem}
\setkomafont{disposition}{\normalcolor\rmfamily}
\setkomafont{caption}{\small\mdseries\selectfont}

\setcapindent{0mm}%
\DeclareMathAlphabet{\mathcal}{OMS}{cmsy}{m}{n}
\usepackage[latin1]{inputenc}
\usepackage[]{SIunits}
\newcommand{\Fig}[1]{Fig.~\ref{#1}}

\newcommand{\wb} {RNDB}
\newcommand{\pgs} {PGS}
\newcommand{\Tb} {T}
\newcommand{\twc} {\tau_w}
\newcommand{\nmod} {\Delta}

\begin{document}
\title{\huge Analysis of transverse Anderson localization in refractive index structures with customized random potential}
\author{\large Martin~Boguslawski,$^{1,*}$ Sebastian~Brake,$^1$ Julien~Armijo,$^{1,2}$ Falko~Diebel,$^1$ Patrick~Rose,$^1$ and Cornelia~Denz,$^{1}$}
\affil{
\normalsize
$^1$Institut f\"ur Angewandte Physik and Center for Nonlinear Science (CeNoS), \\Westf\"alische Wilhelms-Universit\"at M\"unster, 48149 M\"unster, Germany\\
$^2$Departamento de F\'isica, {MSI-Nucleus on Advanced Optics, and Center for Optics and Photonics (CEFOP), }Facultad de Ciencias, Universidad de Chile, Santiago, Chile
\\
$^*$\textcolor{blue}{martin.boguslawski@uni-muenster.de}
}
\date{}
\maketitle
\begin{abstract}
\small We present a method to demonstrate Anderson localization in an optically induced randomized potential. By usage of computer controlled spatial light modulators, we are able to implement fully randomized nondiffracting beams of variable structural size in order to control the modulation length (photonic grain size) as well as the depth (disorder strength) of a random potential induced in a photorefractive crystal. In particular, we quantitatively analyze the localization length of light depending on these two parameters and find that they are crucial influencing factors on the propagation behavior leading to variably strong localization. Thus, we corroborate that transverse light localization in a random refractive index landscape strongly depends on the character of the potential, allowing for a flexible regulation of the localization strength by adapting the optical induction configuration.
\end{abstract}
\section{Introduction}
%%%%%%%%%%%%%%%%%%%%
As a well known but nonetheless fascinating effect in condensed-matter physics, Anderson localization (AL) \cite{Anderson} describes the increased probability of a wave function to be localized in a randomized potential in the vicinity of its initial position. 
As a consequence of disorder, the transport of waves is suppressed and the conduction is diminished causing materials to become insulating. 
Historically, this intriguing effect was investigated primarily in solid-state matters with respect to its electrical conductivity \cite{Lee}.

However, since AL is essentially a coherent wave phenomenon relying on multiple interference effects, it can be observed with any kind of waves, be they matter waves, like electrons in solids, or ultracold atoms \cite{Kondov, Jendrzejewski}, or waves like sound or light \cite{Hu, Wiersma}. 
In the past 20 years the question arose whether AL can be observed in photonic systems as well, featuring a random refractive index potential \cite{John}. 
Optical systems additionally offer the advantage that the non-interacting, linear regime is easily achieved. 
Against this background various photonic systems were employed to successfully demonstrate AL \cite{DeRaedt, Chabanov} in recent years. 
In general, this process helps to establish a deeper insight into the still not completeley understood physics of AL. 
Especially considering photonic random systems, a lot of theoretical discussions were carried out to identify AL in numerous configurations and with respect to various parameters \cite{Feng, Dimitropoulos, Jovic1, Lobanov, Abouraddy}.

Furthermore, important experimental work was done to demonstrate AL of light waves, for instance in a granular medium of high refractive index modulation in three dimensions \cite{Wiersma}.
Transverse localization \cite{DeRaedt} in a medium with a refractive index modulation in one or two dimensions was explored in semi-conductor substrates including nonlinear effects \cite{Lahini} and direct laser written random photonic systems \cite{Rechtsman, Ghosh}, but also in disordered optical glass fibers \cite{Karbasi} as well as in photorefractive crystals \cite{Schwartz, Levi}.

All these works and many more demonstrated that in random (photonic) systems the effect of localization emerges due to the coherent character of wave fields.
In such a system, eigenstates are localized rather than extended, where the exponential decay of the mean envelope of propagating waves is characteristic for AL \cite{Lee}.
Moreover, the last-mentioned contributions introduced the randomization of regular refractive index structures using the optical induction technique, where the authors showed that the regularity-to-randomization ratio can be varied continuously.

% start
%%%%%%%%%%%%%%%%%%%%
In this contriubtion, we stochastically analyze influences of the structural size, which we term as a photonic grain size (\pgs{}), and the disorder strength of optically induced transversely fully random but longitudinally constant refractive index structures on the degree of transverse localization.
To randomize the refractive index of a photorefractive crystal in this manner, we implement a particular light field of transverse randomness and longitudinally uniform intensity, in the following referred to as random nondiffracting beam (\wb{}).
Calculations and corresponding statistics of the \wb{}'s field distribution will be subject of the second section.
Section 3 describes the experimental preparation of random refractive index structures by optical induction, allowing for transverse AL of light. 
Monitoring and analyzing output distributions of probe light fields which show AL is subject of section 4.
In the last section, we explore the influence of a changing random potential on the localization behavior, which is quantified by the localization length.
Therefore, we vary the structural size of the \wb{} in order to analyze the impact of a changing \pgs{} on the localization on the one hand.
On the other hand, we successively increase the disorder strength for a whole stochastic set of potential realizations in order to demonstrate AL of controllable localization strengths.
\section{Random nondiffracting beam}
%%%%%%%%%%%%%%%%%%%%

In general, nondiffracting beams---first studied by Durnin \cite{Durnin1, Durnin2}---are characterized by a modulated intensity distribution transverse to the direction of propagation while the intensity remains constant in longitudinal direction.
As an additional feature, all nondiffracting beams exhibit the intriguing effect of self-healing \cite{Bouchal2}. 
Mathematically, the static field distribution of nondiffracting beams is a solution of the time invariant Helmholtz equation, which is separable into a transverse and a longitudinal part.
The Fourier transform of the transverse field distribution reveals a ring shaped spectrum.
This circle distribution is a condition for the translation invariance in direction of propagation $z$, meaning a longitudinal $k_z$ component of equal length for all contributing wave components.
In turn, the radius of the circle {$k_r$ }determines the structural size of the transverse modulation {$g = \pi/k_r$ }in real space which is highly versatile for various spectra ranging from periodic and quasiperiodic to circular, elliptical and parabolic symmetries \cite{Boguslawski, Bouchal1, Bandres}.
Of course, experimentally implemented nondiffracting beams are not infinitely extended, in fact implying a thin ring rather than a circle which contains all the spatial spectrum.

\begin{figure}[t]
  \center
	\includegraphics[width=.8\textwidth]{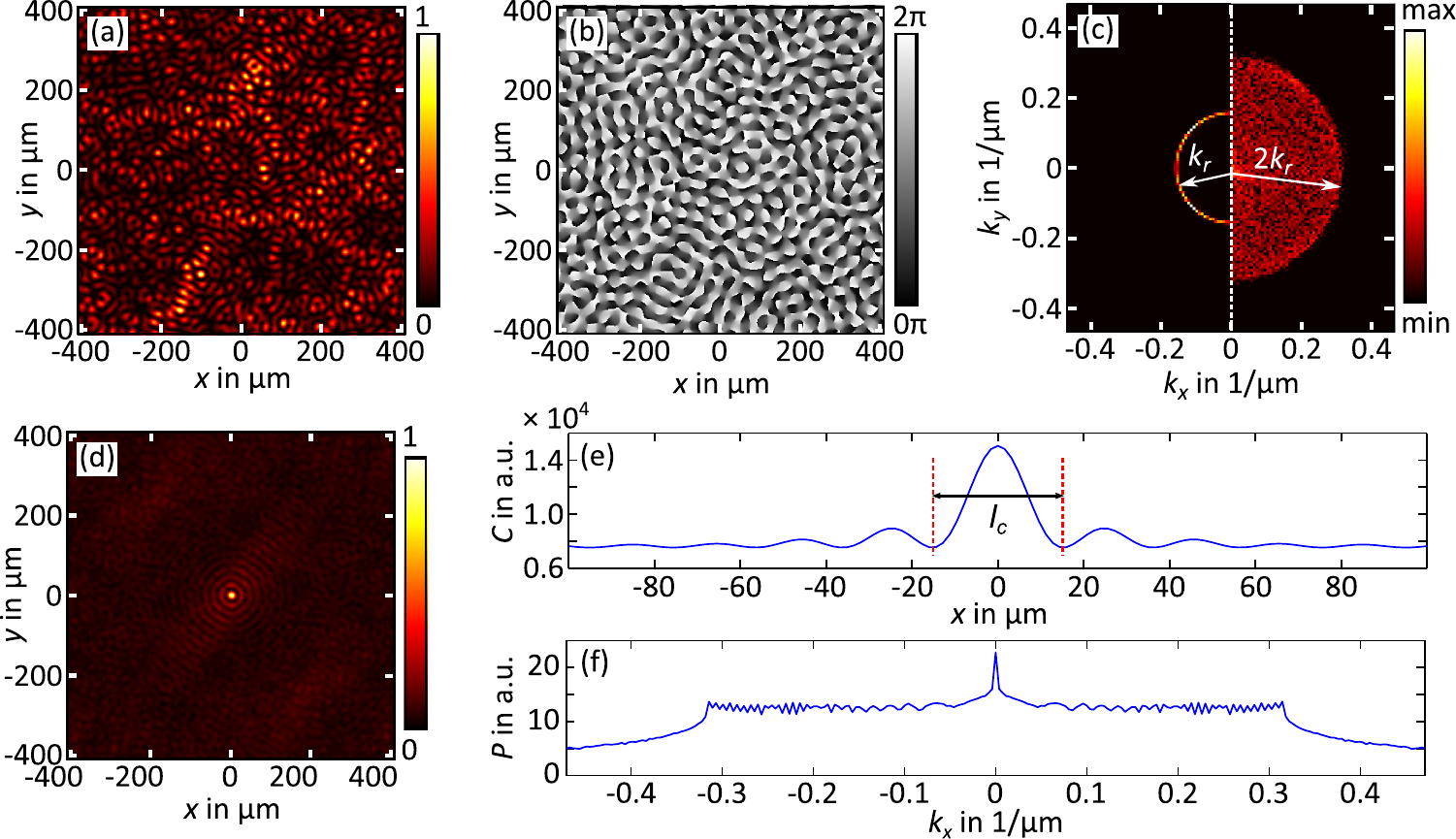}
	\caption{
	Simulated distributions for a randomized nondiffracting writing beam configuration with a \pgs{} of $g = \unit{20}{\micro\meter}$. 
	In (a) the intensity and in (b) the phase distributions in real space are depicted. 
	In (c) a comparison between the spatial spectrum of the writing beam (lhs) and the logarithmic spectrum of its intensity (rhs) is drawn, $k_r$ is the ring radius of the writing beam's transverse Fourier spectrum. 
	Distribution in (d) presents the autocorrelation function of the intensity presented in (a). 
	In (e) the mean autocorrelation through the maximum as well as the corresponding correlation length $l_c$ and in (f) the mean power spectrum through the spectral center is plotted (both averaged over 100 different distributions).
	}
	\label{fig:randNDB}
\end{figure}

We exploit the approach of a ring shaped spatial spectrum to develop a nondiffracting beam of random transverse intensity distribution.
To calculate the field distribution of such a \wb{}, we perturb the phase distribution of the ring shaped spectrum by random values in the interval $[0, 2\pi[$. 
An inverse Fourier transform yields the corresponding field distribution of the random nondiffracting beam in real space \cite{Cottrell}. 
Figure \ref{fig:randNDB} presents one of the calculated distributions in real as well as in Fourier space. 
In real space, the intensity distribution shown in \Fig{fig:randNDB}(a) resembles a random pattern with a particular \pgs{} of $g = \unit{20}{\micro\meter}$. 
This randomness is also apparent in the real space phase distribution presented in \Fig{fig:randNDB}(b). 
The power spectral density $P$, depicted logarithmically in \Fig{fig:randNDB}(c, rhs), represents a lowpass filtered spectrum with a sharp frequency limit. 
This limit is determined by the \pgs{} of the \wb{}'s intensity and is given by the double radius of the Fourier ring of the \wb{} as shown in \Fig{fig:randNDB}(c, lhs). 
Additionally we observe an accumulation of the spectral weighting on the outside circle of the power spectrum $P$ and near its center, which is depicted in Figs. \ref{fig:randNDB}(c, rhs) and {\ref{fig:randNDB}}(f). 
We further characterize the intensity by the 2D autocorrelation $C(\vec r)$ function of the intensity. 
Therefore, we use the Wiener-Khinchin theorem to determine $C$.

The resulting autocorrelation function features a central intense lobe, whose width $l_c$, defined as the diameter of the first dark ring, is the correlation length of the disorder and scales directly on the chosen \pgs{}. 
Slices of the autocorrelation function and the power spectral density both averaged over 100 different random intensity distributions are shown in \Fig{fig:randNDB}(e) and {\ref{fig:randNDB}}(f). 
In the former, one clearly can notice a series of local maxima beyond the central lobe. 
Their presence is due to the particular disorder we use, with a spectrum contained in a disk with more weight on the outside circle [cf. \Fig{fig:randNDB}(f)].
However, altogether no prominent long-range order is significant for the random intensity structure.

\section{Optical induction of random refractive index distributions}
In analogy to the technique of optical induction of periodic and quasiperiodic as well as curvilinear refractive index modulations \cite{Rose}, we use the introduced \wb{}s to generate a random index modulation. 
Since the propagation behavior in random media will be analyzed stochastically we generate a set of 100 different random structures of fixed \pgs{} which is precalculated and stored to achieve reproducibility of the experiment.
% setup, experimental implementation
%%%%%%%%%%%%%%%%%%%%%%%%%%%%%%%%%%%%%%%%

\begin{figure}[b]
  \center
	\includegraphics[]{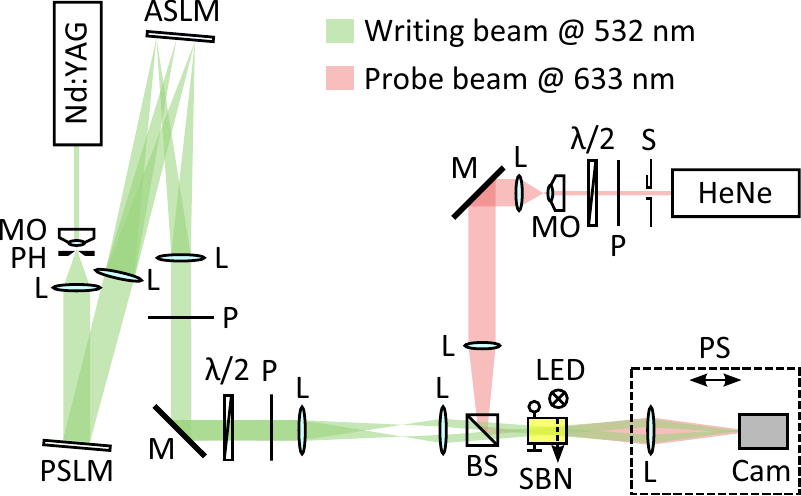}
	\caption{Sketch of the experimental setup to induce random photonic structures. A/PSLM: amplitude/phase spatial light modulator, BS: beam splitter, Cam: camera, L: lens, LED: white light emitting diode, $\lambda/2$: half-wave plate, M: mirror, MO: microscope objective, P: polarizer, PH: pin hole, PS: positioning stage, S: shutter, SBN: photorefractive crystal.}
	\label{fig:figure2}
\end{figure}

In order to induce the random 2D refractive index landscape, we use the experimental setup presented in \Fig{fig:figure2}. 
Employing a frequency doubled solid state laser (Nd:YAG, \unit{\lambda_w = 532}{\nano\meter}), we modulate the wave field of a plane wave with a phase-only spatial light modulator (PSLM). 
For these experiments we solely randomize the phase of the spatial spectrum by introducing an appropriate random phase distribution to the PSLM.
The modulation of the phase is sufficient to reproduce the desired intensity distribution since an adequate propagation distance transfers the phase modulation to an according amplitude modulation of the field. 
In this context we use an amplitude modulator (ASLM) and a linear polarizer (P) to filter the relevant spatial frequencies in Fourier space.
A half-wave plate ($\lambda/2$) and a second polarizer provide the proper linear polarization for the writing configuration (see further below for details).

Subsequently, a 4$f$-arrangement of two lenses images a \wb{} into a volume of interest where the crystal is placed.
To guarantee a distribution of refractive index invariant in the direction of propagation, the interference volume of the writing beam has to be analyzed. 
We implement a computer controlled positioning stage with an imaging system including a camera and a lens which enables a translation over $\unit{100}{\milli\meter}$ in the direction of propagation. 
In this manner we are able to record the transverse intensity distribution in particular planes, for instance to detect the intensity of the writing beam at different propagation lengths. 
By stacking all planes we get the full three-dimensional intensity information. 

\begin{figure}[t]
  \center
	\includegraphics[]{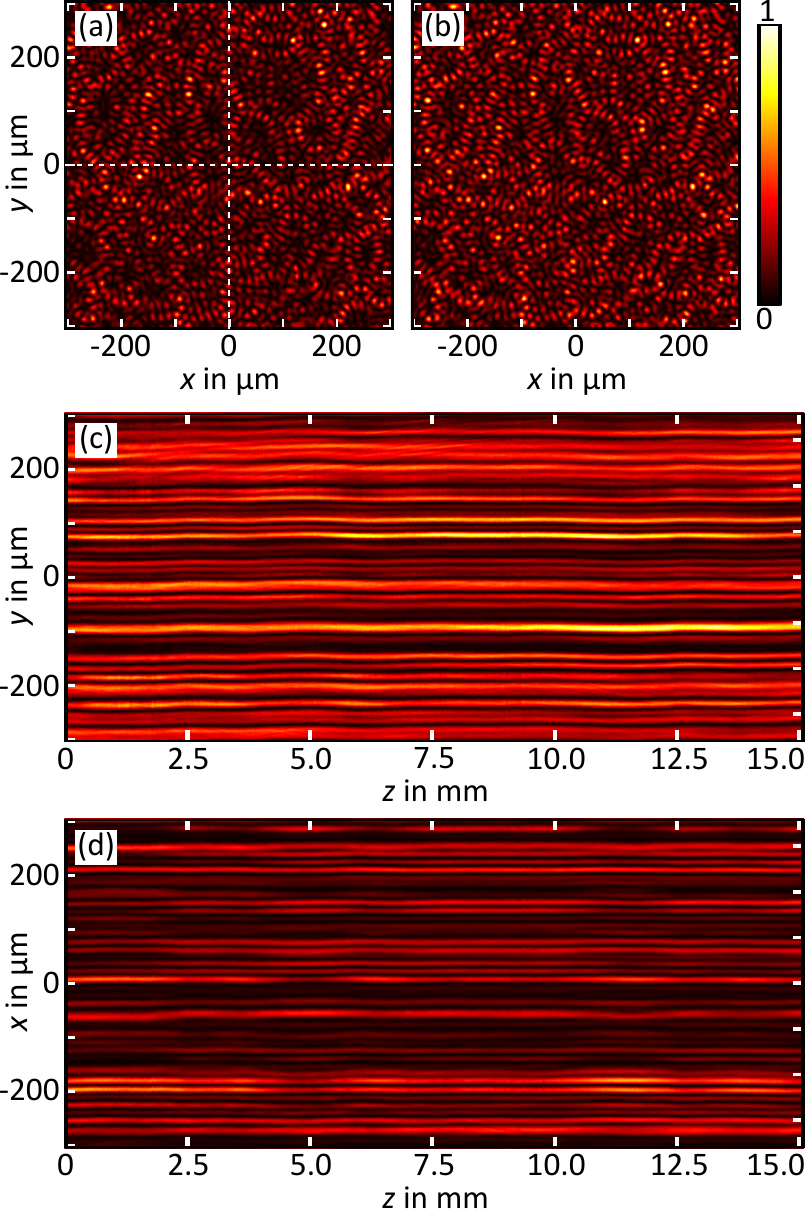}
	\caption{
	Experimentally recorded intensity profiles of a nondiffracting writing beam with random transverse intensity distribution. 
	(a) and (b) present nearly identical transverse intensity modulations at the front and the back face of the crystal, (c) and (d) show the intensity development with propagation in $z$ direction in the central vertical and horizontal longitudinal plane, cf. dotted lines in (a).}
	\label{fig:figure3}
\end{figure}

In \Fig{fig:figure3} we present the different intensity distributions in transverse as well as in longitudinal planes. 
A comparison between the transverse intensity distribution at the front face with the one at the back face can be drawn in Figs. \ref{fig:figure3}(a{) and \ref{fig:figure3}(}b). 
These transverse and longitudinal intensity distributions indicate that nondiffracting field distributions with random transverse intensity modulation are very well suited for being used as writing beams in order to optically induce longitudinally elongated 2D random photonic structures.

For the optical induction of a refractive index structure, the photorefractive medium, namely a cerium doped strontium barium niobate (SBN) crystal, is illuminated with a corresponding light field while externally biased.
In general, the SBN crystal exhibits an anisotropic electrooptic characteristic \cite{Terhalle1}.
That is, for the extraordinary polarization (parallel to the crystal's symmetry axis, $c$ axis), the relevant electrooptic coefficient $\unit{r_{33} = 235}{\pico\meter\per\volt}$ is approximately five times larger than the relevant one for an ordinarily polarized wave field, $\unit{r_{13} = 47}{\pico\meter\per\volt}$ \cite{Vazquez}. 
Thus the effect of the induced refractive index on an extraordinarily polarized light field is significantly higher compared to a field holding ordinary polarization.

Applied to our optical system, this anisotropy allows us to switch between writing and probe beam propagation: 
In first approximation, the writing beam does not experience the induced structure, while for the probe beam which will be introduced in the next section, the refractive index change is proportional to the writing beam intensity.

\section{Transverse localization in random structures}
During the probing process of the written structure, the external field is switched off and---provided that the intensity of the probe beam is small---the modulation of the refractive index remains until active deletion. 
That is, the refractive index can be re-homogenized by illuminating the SBN sample for several seconds with a bright incoherent cold-light source.

\begin{figure}[b]
  \center
	\includegraphics[width = .9\textwidth]{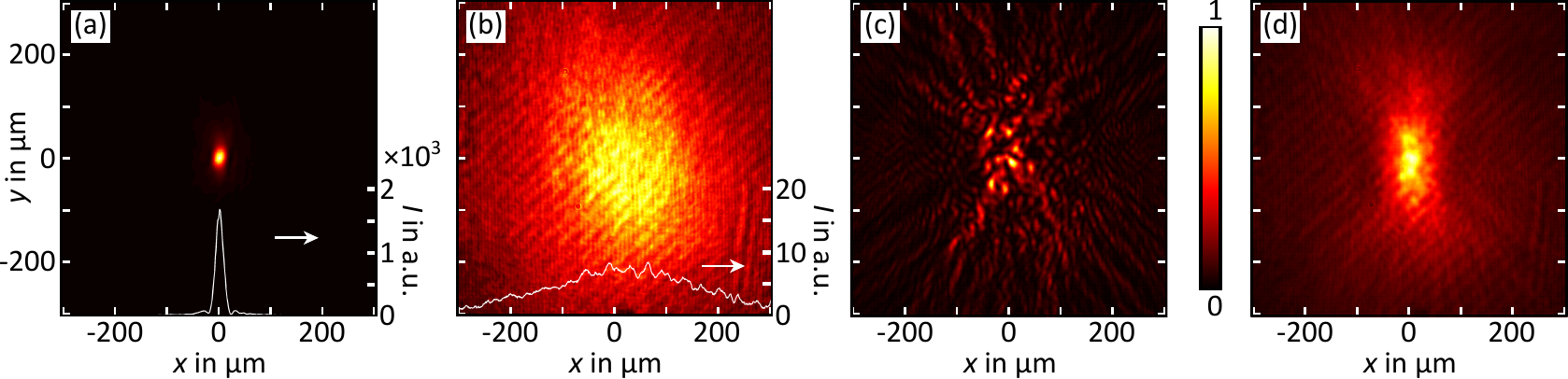}
	\caption{Intensity distribution of probe beam in unmodulated medium at (a) crystal's front face, (b) back face; in randomly modulated medium at back face: (c) single shot for particular random potential, (d) mean intensity averaging over a set of 100 single probe shots. All intensities are normalized.}
	\label{fig:figure4}
\end{figure}
We introduce the probe beam as a tightly focused Gaussian beam of a HeNe laser with $\lambda_p = \unit{633}{\nano\meter}$.
In \Fig{fig:figure2}, the probe beam setup is depicted as the red arm where a combination of lenses provides a tight focus and a polarizer and a half-wave plate again configure the proper linear polarization state parallel to the $c$ axis of the SBN crystal.
The Gaussian beam waist is directly positioned in front of the crystal and its transverse position defines the input center, as depicted in \Fig{fig:figure4}(a). 
Due to the tight focusing to $w_0 = \unit{(17.8\pm0.7)}{\micro\meter}$ Gaussian beam waist, the spatial spectrum of the probe beam is broad which enables an analysis of the transmission behavior through the random potential landscape for various spatial frequencies.

\begin{figure}[t]
  \center
	\includegraphics[width = 0.45\textwidth]{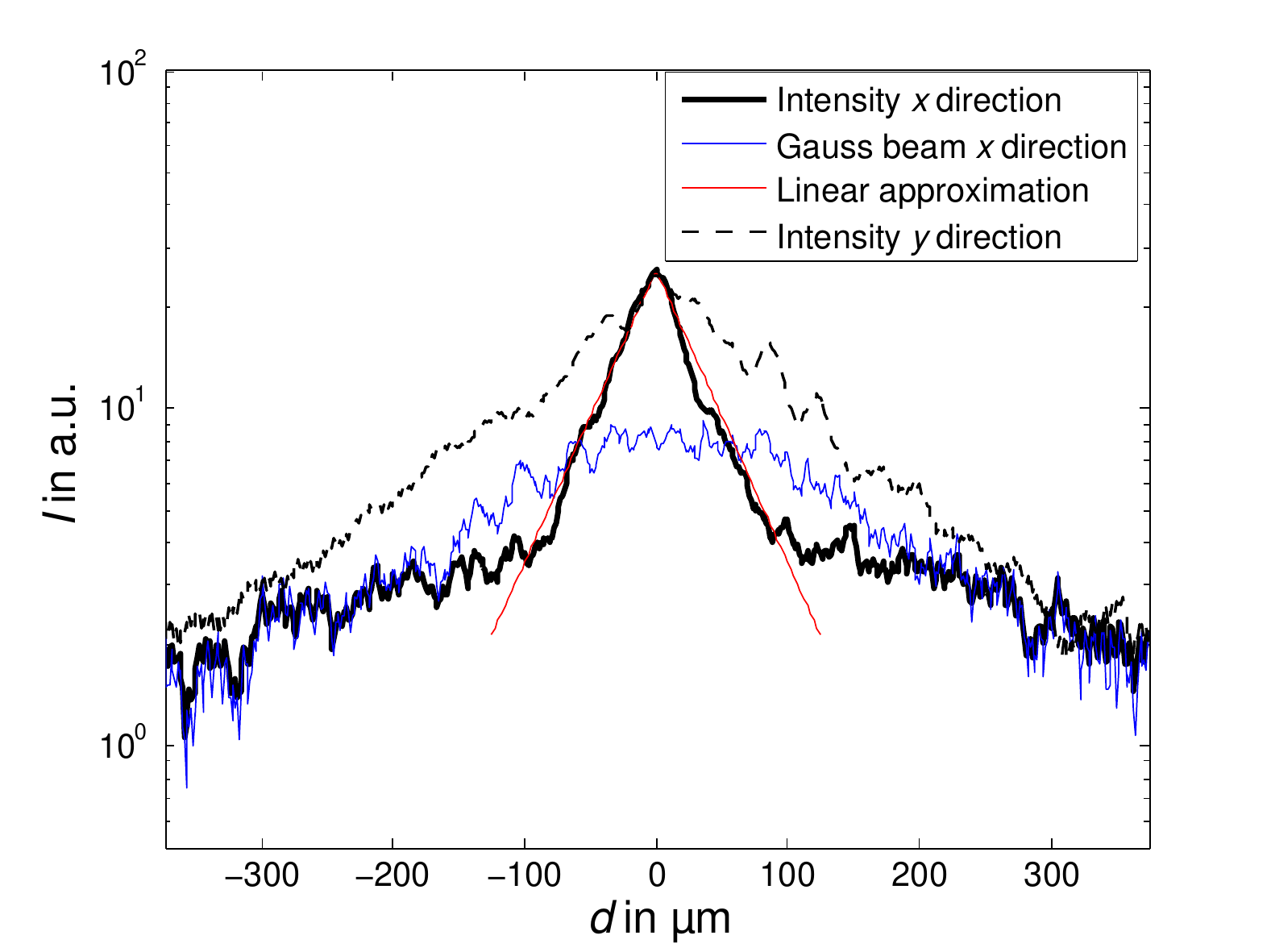}
	\caption{Logarithmic plot of the central intensity $I$ for blue line: potential without disorder, black solid line: strongly modulated potential in $x$ direction. Red lines mark linear profiles of logarithmic intensity in $x$ direction around the localization center and corresponding mean slope. Black dashed line: strongly modulated potential in $y$ direction.}
	\label{fig:figure5}
\end{figure}
In a potential absent of refractive index modulation the probe beam experiences a transverse beam broadening due to diffraction, as shown in \Fig{fig:figure4}(b). 
This configuration describes the lower limiting case of a general behavior where for significant potential modulation the broadening is suppressed due to AL, resembling transversely an exponential distribution in the vicinity of the input position of a wave field \cite{DeRaedt}.

A set of random intensity modulations is characterized by a fixed \pgs{} $g$ and illumination time $\Tb$, leading to comparable contrasts of the induced refractive index modulations.

In order to probe each single potential landscape, the intensity distribution of the Gaussian input probe beam at the back face of the crystal is recorded. 
Exemplarily, a single representative intensity distribution of a probed random potential is depicted in \Fig{fig:figure4}(c). 
Here and in all single probe measurements, the transverse center of the back plane of the crystal corresponds to the input center at the front face of the crystal. 
Thus, the Gaussian probe beam's direction of propagation is perpendicular to each transverse plane showing refractive index modulation.

Subsequently, we analyze stochastically the intensity distribution of the whole set by calculating the average of the transverse intensity at the output face of the crystal, as shown in \Fig{fig:figure4}(d) for $g = \unit{16.1}{\micro\meter}$. 
To identify localization, we show a slice of the 2D intensity profile along the $x$ direction, i.e. an intensity profile integrated over 11 lines in \Fig{fig:figure5}. 
One clearly identifies a region around the central position where the intensity distribution decays exponentially. 
This is indicated by a linear slope in the logarithmic scale which is a prominent indication of AL. 
The localization lenght $\xi$ is obtained by fitting the $x$ profile of the localized intensity $I(\vec r)$ by the behavior:
\begin{eqnarray}
I({x}) = \exp\left(-2\frac{\left|{x} - {x_0}\right|}{\xi}\right).
\end{eqnarray}
Here $x_0$ indicates the beam center. For a particular set of random structures with $g = \unit{16.1}{\micro\meter}$, we find $\xi = \unit{{(100.2\pm 1.1)}}{\micro\meter}$ as shown in \Fig{fig:figure5}. The uncertainty is given through the error estimates of the slopes compared to their linear fitting functions. 

In the following considerations we exclusively investigate the localization behavior of light in the direction parallel to the $c$ axis. 
Since a significant orientational anisotropy is specific for SBN crystals due to a strong drift of the charge carriers in direction of the external field \cite{Terhalle1}, the modulation of the refractive index and connected to that the disorder strength parallel to the $c$ axis are much stronger than perpendicular to the symmetry axis. 
This causes an asymmetry of localization in these two directions where the localization in vertical direction is much weaker than parallel to the $c$ axis. 
For comparisons of both orientations we added the logarithmic intensity plot in vertical direction in \Fig{fig:figure5} (dashed black line) and find that the localization is much weaker in this direction.

To clearly identify the localization behavior in a random potential in contrast to the case without randomization, the output intensity for a vanishing disorder strength is added in \Fig{fig:figure5}. 
This profile is roughly parabolic, since the intensity distribution for this particular case resembles a broadened Gaussian function due to the influence of diffraction on the input light field configuration.

Comparing its distribution with the localized one, it becomes obvious that the localization effect does not occur all over the recorded area but up to a finite distance from the input center. 
Moreover, the outer tails of the distribution of the random case are comparable to the Gaussian beam. 
For an intermediate region, the localized case is less brighter than the Gaussian, which of course is a consequence of energy conservation. 
This behavior identifies a noteworthy characteristic---an amount around a determined transverse distance from the injection center rather than the complete contributing probe field is localized.

\section{Photonic grain size and degree of randomness}

\begin{figure}[t]
  \center
	\includegraphics[width = 0.45\textwidth]{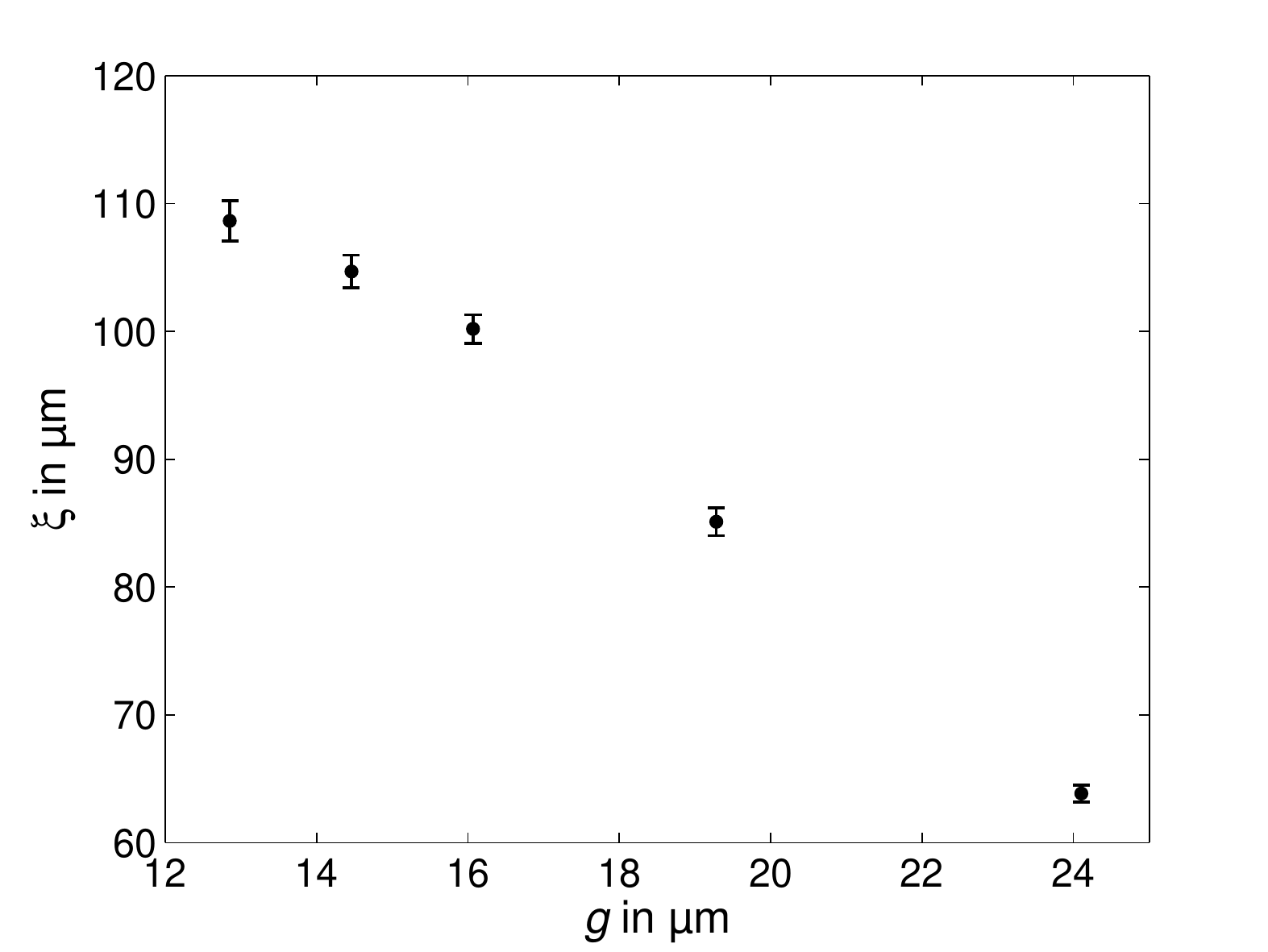}
	\caption{
	Dependency of localization length $\xi$ against $g$. 
	Error determined by linear approximation of localization length.
	}
	\label{fig:figure6}
\end{figure}

Since we developed a setup which enables in a highly flexible manner the induction of a random potential providing the conditions for AL, we are now able to explore the influence of the shape of the refractive index on the localization behavior.
In the following we will thus focus on the effect of the \pgs{} as well as the disorder strength of the random structure, since these are two major properties of the underlying potential determining the propagation characteristics of light.

We firstly vary the \pgs{} by changing the structural size of the set of \wb{} from $g = \unit{13}{\micro\meter}$ to $\unit{24}{\micro\meter}$. 
In \Fig{fig:figure6} we plot the localization length $\xi$ against the structural size $g$. 
Here, an inverse relationship is prominent which shows that with increasing structural size the localization length is decreased.
Thus, the \pgs{} of the random potential has a crucial effect on the localization length, where for larger \pgs{} the propagating light experiences a stronger localization compared to modulation of the refractive index on smaller scales.
{Considering a photonic waveguide system, one explanation for such a dependency could be the enhanced coupling between adjacent refractive index maxima when reducing the mutual distance which is directly connected to the \pgs{}.
However, a detailed explanation of this behavior would need further investigations to identify the essential cause.}

We further investigate the {modulation depth} of the potential as an influencing parameter of the localization length.
{In this context, the modulation depth can be considered as a disorder strength $\nmod$ of the system}.
To control {this parameter} we adapt the illumination time as the refractive index modulation develops exponentially with enduring illumination \cite{Maniloff}.

\begin{figure}[b]
  \center
	\includegraphics[width = 0.45\textwidth]{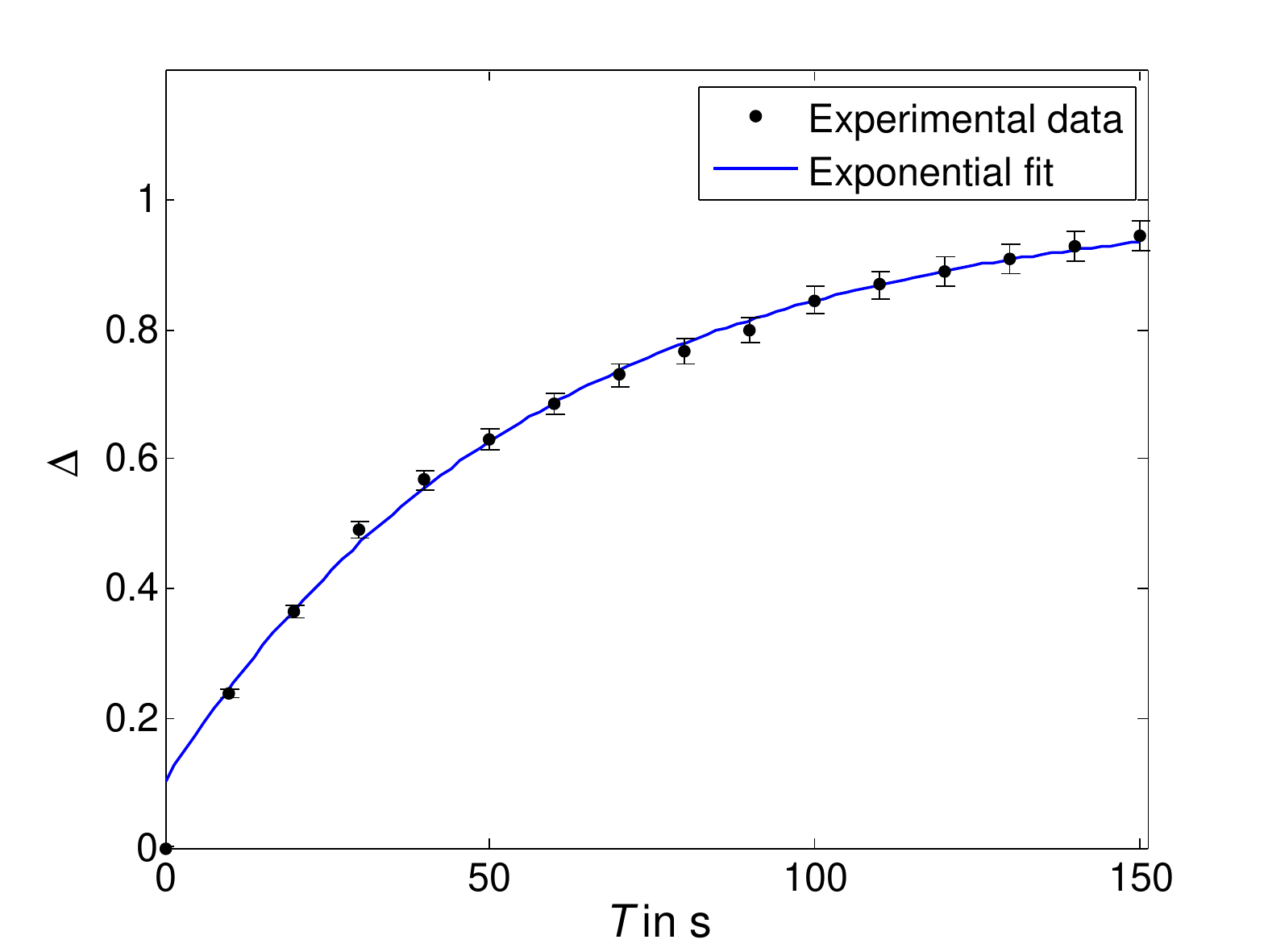}
	\caption{
	Plot of the {disorder strength $\nmod$ with $g = \unit{16.1}{\micro\meter}$} against the illumination time and exponential fit (blue line).
	}
	\label{fig:figure7}
\end{figure}
In order to extract a measure for the {disorder strength} as a function of the illumination time we employ the technique of waveguiding \cite{Terhalle2}. 
Thus, during the illumination sequence we send a plane wave onto the random potential landscape of particular contrast and record the intensity distribution at the back face of the crystal. 
Expecting that every change of the intensity is caused by an increased refractive index contrast, we sum up every absolute intensity change of each pixel after one illumination period. 
According to that, we calculate the mean value over all pixels. 

{Since we assume that the disorder strength $\nmod$ is governed by the refractive index modulation depth with exponential relation on the illumination time, we expect an exponential behavior for $\nmod$:}
\begin{eqnarray}
\nmod({\Tb}) = 1 - \exp(-{\Tb}/\twc),
\label{eq:tempRI_developmt}
\end{eqnarray}
where the temporal development of $\nmod$ depends on the time constant $\twc$.

\begin{figure}[t]
  \center
  \includegraphics[width = 0.45\textwidth]{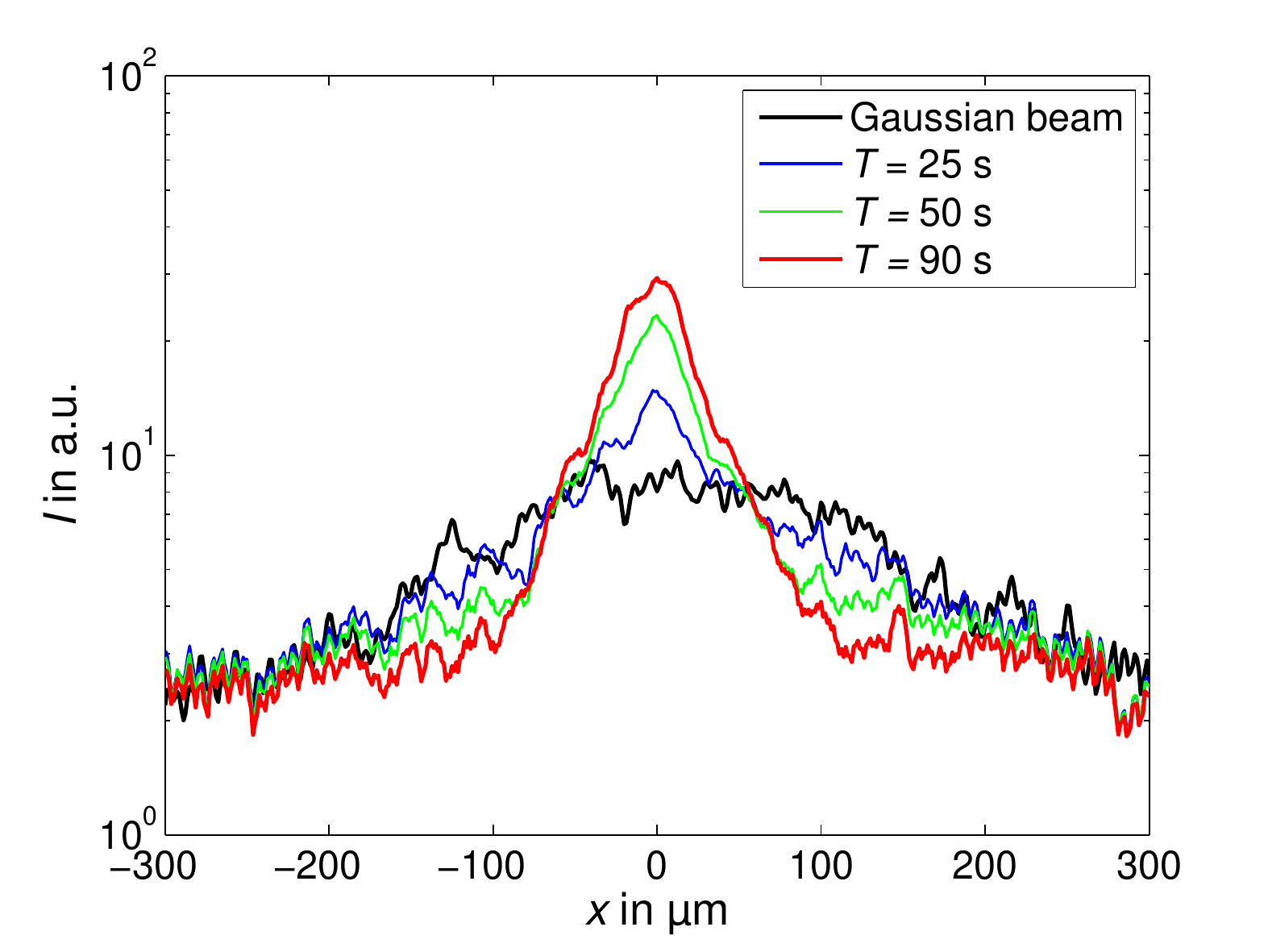}
	\caption{Logarithmic plot of the central intensity $I$ for random potential of different illumination times {and $g = \unit{16.1}{\micro\meter}$}, black line: $\Tb = \unit{0}{\second}$, blue line: $\Tb = \unit{25}{\second}$, green line: $\Tb = \unit{50}{\second}$, red line: $\Tb = \unit{90}{\second}$.}
	\label{fig:figure8}
\end{figure}

\begin{figure}[b]
  \center
	\includegraphics[width = 0.45\textwidth]{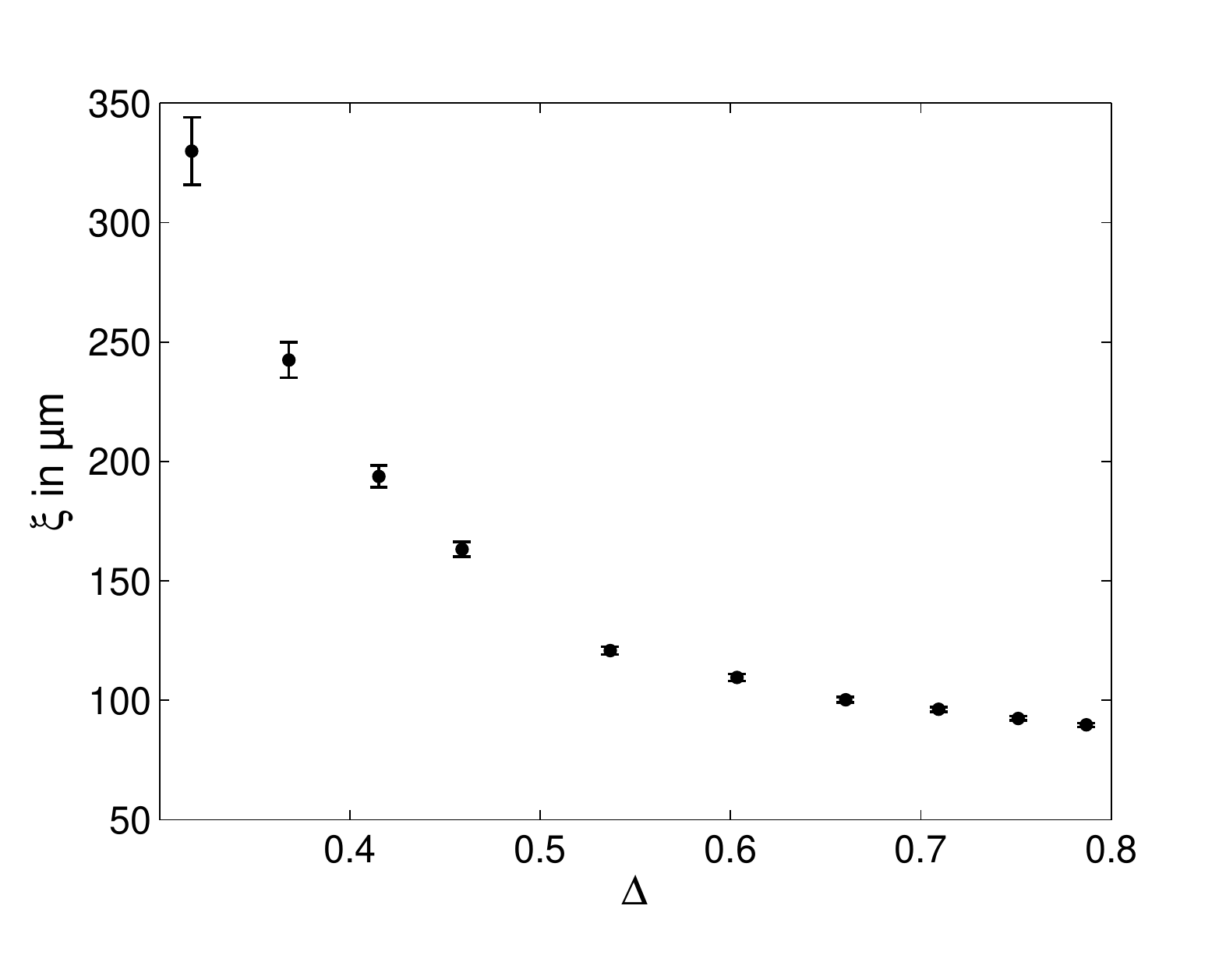}
	\caption{Relation of localization length $\xi$ and relative disorder strength $\nmod$ {for $g = \unit{16.1}{\micro\meter}$}.}
	\label{fig:figure9}
\end{figure}

In \Fig{fig:figure7} the disorder strength $\nmod$ {is plotted }against the illumination time $\Tb$. 
{Although there is a mismatch between data and fitting model for short illumination times, the exponential model shows sufficient accuracy for longer illumination. 
In particular, data points for larger $\Tb$ and therefore for stronger disorder are more relevant for our investigation since the localization length $\xi$ can be determined more precisely (cf. \Fig{fig:figure9}).}
{Alltogether}, approximating the measured data enables us to determine the relative {disorder} strength as a function of illumination time $\Tb$, obtaining the respective time constant $\twc = \unit{(64.0{\pm0.4})}{\second}$.
{Here the error is the standard deviation obtained through the analysis of a set of five pictures per illumination time}.

To {further} identify a dependency of the localization length on the strength of disorder $\nmod$, we successively induce a set of 100 random photonic structures of $g = \unit{16.1}{\micro\meter}$ with various writing times. 
This can be achieved by changing between writing and probing the photonic structure after particular illumination times $\Tb$ {where the structure is probed each $\unit{5}{\second}$ until $\Tb = \unit{30}{\second}$ and each $\unit{10}{\second}$} up to a maximal illumination time of $\unit{90}{\second}$.
Comparing with \Fig{fig:figure7}, such an illumination corresponds to an $80 \%$ saturated refractive index structure. 
In \Fig{fig:figure8} localization profiles for three different illumination times and for the initial state are plotted logarithmically. 
Here, a successive change of the central distribution is significant---from short illumination times resembling a weak localization area to a pronounced region of localization showing an increasingly steep linear slope of logarithmic intensity. 
Plotting the respective localization lengths $\xi$ against the strength of disorder $\nmod$ as presented in \Fig{fig:figure9}, a monotonically decreasing dependency can be extracted for increasing disorder. 
Again, one can affirm that for a higher disorder the coupling between adjacent refractive index maxima is diminished resulting in stronger AL.
This in turn implies, by adapting the writing time of the index modulation in order to vary the disorder strength of the potential we are able to control the localization length in a highly adaptable way.

\section{Conclusion}
To conclude, we developed a method to randomly modulate the refractive index of a photorefractive crystal implementing an optical induction method by use of nondiffracting writing beams with random transverse intensity distributions. 
We found that such a photonic structure exhibits Anderson localization, since a Gaussian input distribution localizes exponentially around the input center after propagation in the random potential. 
Further we quantitatively investigated on parameters influencing the localization length where the structural modulation size identified as a photonic grain size as well as the disorder strength of the refractive index allow for manipulating the localization ability of the photonic system. 
In this context, we further presented an alternative way to Anderson localization by successively increasing the disorder strength of a random potential rather than to increase the order-to-disorder ratio in an initially regular pattern. 
Thus, the introduced model system enabling to adapt the photonic random potential is highly suitable to be applied for the investigation on light localization in potentials of changing random conditions.

\section*{Acknowledgments}
We acknowledge support by Deutsche Forschungsgemeinschaft and Open Access Publication Fund of University of Muenster. {J. A. acknowledges support from Programa ICM P10-030-F and Programa de Financiamiento Basal de CONICYT (FB0824/2008).}

\end{document}